\documentclass[epjc,preprint]{revtex4}
\usepackage{amsfonts}
\usepackage{amsmath}
\usepackage{amssymb}
\usepackage{graphicx}
\usepackage{float}
\usepackage{bm}
\usepackage{caption}
\usepackage{subcaption}
\usepackage{booktabs}
\usepackage{textcomp}
\usepackage[colorlinks=true,citecolor=blue,urlcolor=cyan,linkcolor=blue]{hyperref}

\begin{document}

\title{Lunar tidal response-induced clock frequency variations on the Moon}
\author{Hongbin Zhang}
\author{Baocheng Zhang}
\email{zhangbaocheng@cug.edu.cn}
\affiliation{School of Mathematics and Physics, China University of Geosciences, Wuhan
430074, China}

\begin{abstract}
As clock performance improves, the instability and uncertainty of the
clock's fractional frequency shift (FFS) become comparable to the FFS caused
by the lunar tidal response (LTR), so the influence of the LTR on lunar time
needs to be investigated. In this paper, we used the Keplerian expansions
method to study the FFS to the level of $10^{-18}$ caused by the LTR between
two clocks placed in different positions on the Moon. We consider the case
of the Moon orbiting in an elliptical orbit, which is closer to the actual
situation compared to the case of the Moon moving in a circular orbit. An
interesting phenomenon is found that the longitude difference between two
clocks can cause a change in the amplitude and phase of the FFS due to the
LTR, but the latitude difference only changes the amplitude of the FFS. We
further estimated the clock difference between two clocks near the Moon
caused by the LTR. When one clock is on the lunar surface, and the other is
on the Earth's surface, the LTR will cause a clock difference of $%
-5.68\times 10^{-7}$ $\mu $s/day between the two clocks. These are
significant for the construction of a lunar time and clock synchronization
near the Moon.

\end{abstract}

\maketitle

\section{Introduction}

A multinational group, including NASA, is preparing for a return to the Moon
under the Artemis Accords \cite{artemis20,ESA,NASA}. China is also planning
to land on the Moon around 2030 \cite{china} with many exciting results of
the lunar exploration project being announced one after another \cite%
{che2021age,tian2021non,cui2024sample,wang2024scientific}. As assets on the
lunar surface are expected to increase significantly in the near future,
precise positioning, navigation, and time synchronization of lunar surface
devices will become issues of greatest interest.

Recently, the International Astronomical Union (IAU) has approved a proposal
to establish a standard lunar celestial reference system and lunar
coordinate time \cite{soffel2003iau,kaplan2006iau}. Several methods for
defining lunar time have been proposed \cite%
{kopeikin2024lunar,ashby2024relativistic}, where the proper time measured by
clocks at different locations on the lunar surface is affected by both
gravitational and kinematic effects \cite{turyshev2025relativistic}. These
effects are essential for constructing a consistent and high-precision lunar
time system.

Previous studies have investigated relativistic frequency shifts and clock
comparisons in gravitational fields. For example, Vessot et al. \cite%
{vessot1980test} measured gravitational and kinematic frequency shifts at
the order of $(c^{-2})$, while Linet et al. \cite{linet2002time} extended
the theoretical framework to rotating bodies. Wolf et al. \cite%
{wolf1995relativistic} evaluated the contribution of tidal potentials to
clock frequency shifts, and Ref. \cite%
{qin2020tidal,qin2023tidal,zhang2026frequency} analyzed fractional frequency
shifts (FFS) between clocks at different locations on the Earth caused by
tidal effects. More recently, Zhang et al. \cite{zhang2025frequency} studied
FFS between Earth-based and lunar surface clocks, including both tidal and
non-tidal contributions, and Ashby et al. \cite%
{fermi1922phenomena,ashby1986relativistic,ashby2024relativistic} developed a
relativistic framework for clock rates on the Moon, later extended by
Kopeikin et al. \cite{kopeikin2024lunar} to include tidal potentials.

However, most existing studies either focus on the Earth or treat the Moon's
orbit as circular \cite%
{williams2015tides,williams2014lunar,zhang2026frequency}, and the role of
the lunar tidal response (LTR) in realistic orbital configurations has not
been systematically investigated. In particular, the impact of LTR on
high-precision clock comparison and lunar time construction in an elliptical
orbit remains insufficiently explored.

Tidal response is a fundamental phenomenon not only in planetary systems but
also in broader astrophysical contexts. For instance, LTR can be used to
infer the thermal structure of the Moon's interior \cite{Park2025}, while
tidal responses in other bodies, such as Io \cite{Park20251} and even black
holes \cite{Bhatt2024,Katagiri2025}, provide insights into their internal or
spacetime properties. These studies highlight the broader significance of
tidal effects and motivate their inclusion in precision timing models.

With the rapid development of atomic clock technology, the instability and
uncertainty of clocks have reached the $(10^{-18})$ level or better \cite%
{bloom2014optical,nicholson2015systematic,huntemann2016single,nakamura2020coherent,boulder2021frequency,marshall2025high}%
. At this level of precision, the FFS induced by LTR becomes comparable to
the intrinsic uncertainties of state-of-the-art clocks. Therefore, LTR must
be taken into account in high-precision clock comparison experiments near
the Moon.

In this work, we introduce the LTR into the Lunar Celestial Reference System
(LCRS) and analyze its effect on clock frequency shifts. We model the LTR in
an elliptical orbit using the Keplerian expansion method, providing a more
realistic description than circular-orbit approximations. We incorporate the
tidal response consistently into the relativistic framework of LCRS and
quantify the impact of LTR on FFS and the time difference between clocks at
different locations on and near the lunar surface. Our results show that
this effect is not negligible for next-generation high-precision timing
applications.

This paper is organized as follows. In Sec. II, we derived the tidal
potential and tidal response under the Kepler expansion method. We then
introduced the LTR into the LCRS. By estimating the magnitude of different
factors affecting the FFS, we found that it is interesting and worthwhile to
focus on the LTR, which cannot be ignored for future construction of lunar
time and the synchronization of clocks near the Moon. In Sec. III, we
obtained the FFS between two clocks at different locations on the lunar
surface. In Sec. IV, we will analyze the time difference between two clocks
near the Moon caused by the Moon's tidal response. Finally, we give the
conclusion in Sec. V. The symbols used in this paper were explained upon
their introduction and are also summarized in Table 1.

\begin{table}[tbp]
\caption{Notations used in this paper.}
\label{tab:const1}\centering
\renewcommand{\arraystretch}{1}
\begin{tabular}{cp{12cm}}
\hline\hline
Lunar orbit parameters &  \\ \hline
$e$,$a$,$T$,$\omega$,$\Omega$,$\epsilon$ & The eccentricity of the lunar
orbit, the semi-major axis, the true anomaly, the argument of perigee, the
longitude of the ascending node, and the orbital inclination. \\
$L$,$M$ & The mean longitude and mean anomaly of the lunar orbit. \\ \hline
Angular arguments &  \\ \hline
$\phi$,$\lambda$ & The latitude and longitude of the clock. \\
$\psi$,$\psi_s$ & The angles between the clock's position vector and the
position vectors of the Earth and the Sun, respectively, within the LCRS. \\
\hline
Sign conventions &  \\ \hline
$K$,$P_A$,$P_B$ & Function defined by the LCRS metric and functions defined by the
tidal response potential. \\
$C_A$,$C_B$,$D_A$,$D_B$ & Functions defined by the position of the clock and the position of
the Moon. \\ \hline
Subscripts &  \\ \hline
$E$,$m$,$s$ & Label for the Earth, Moon, and Sun. \\
$t$,$tE$,$ts$,$tr$ & Label for the tidal potential,Earth's tidal potential,
the Sun's tidal potential, and tidal response. \\
$mE$,$ms$ & Labels for the distance from Earth to the Moon and the distance
from the Sun to the Moon. \\
$l$,$q$ & Label for the degree and order of the Moon's gravitational
potential. \\
$A$,$B$ & Label for the clock A and clock B. \\ \hline\hline
\end{tabular}%
\end{table}

\section{The local frame of the Moon}

The fundamental assumption of general relativity is that the laws of physics
do not depend on the chosen coordinate system \cite{turyshev2013general}.
This premise is achieved by assuming the invariance of Riemannian spacetime
intervals. Choosing an appropriate coordinate system will make the specific
physical problem being analyzed simpler and the logic clearer. The choice of
the local Moon's center of mass frame is appropriate when we consider that
clocks near the lunar surface are affected by the gravitational fields of
the Earth, the Sun, and the Moon.

In accordance with the recommendations of the IAU resolutions \cite%
{soffel2003iau}, the space-time metric for the LCRS \cite%
{kopeikin2024lunar,Bourgoin} is constructed in a manner analogous to that of
the Geocentric Celestial Reference System (GCRS) metric \cite%
{qin2019relativistic}, wherein Earth-related quantities are replaced by
Moon-related quantities. The metric tensor in the LCRS coordinates ($%
ct,r^{i} $) can be written in the two-potential form
\begin{equation}
\begin{aligned} g_{00}(t, r)= & 1-\frac{2\Phi(t, r)}{c^{2}}+ \frac{2\beta
\Phi^2(t, r)}{c^{4}} +\mathcal{O}\left(c^{-6}\right) \\ g_{0 i}(t, r)= &
\frac{2(\gamma +1)\delta_{i j}\Phi^j(t,
r)}{c^{3}}+\mathcal{O}\left(c^{-5}\right) \\ g_{i j}(t, r)= & -\delta_{i
j}\left[1+\frac{2 \gamma \Phi(t, r)}{
c^{2}}\right]+\mathcal{O}\left(c^{-4}\right).\label{me} \end{aligned}
\end{equation}%
where $c$ is the speed of light, which does not change when transforming
coordinates, $\gamma $ and $\beta $ are the post-Newtonian parameters (in
general relativity, $\gamma =\beta =1$). Here, $\Phi (t,r)$ and $\Phi
^{j}(t,r)$ are the scalar and vector harmonic potentials, respectively,
which generally are written as
\begin{equation}
\Phi (t,r)=\Phi _{m}(t,r)+\Phi _{t}(t,r)+\Phi _{iner}(t,r),
\end{equation}%
and
\begin{equation}
\Phi ^{j}(t,r)=\Phi _{m}^{j}(t,r)+\Phi _{t}^{j}(t,r)+\Phi _{iner}^{j}(t,r).
\end{equation}

From the assumption of the invariance of the Riemannian space-time interval $%
ds^{2}=g_{\mu \nu }dx^{\mu }dx^{\nu }$, the relationship between coordinate
time $t$ and proper time $\tau $ of the clock near the Moon is given by
\begin{equation}
\mathrm{d}\tau =\sqrt{g_{00}+2g_{0i}\frac{v^{i}}{c}+g_{ij}\frac{v^{i}v^{j}}{%
c^{2}}}\mathrm{d}t,  \label{pt1}
\end{equation}%
where $v=dr/dt$ is the speed of clock in the LCRS, and $v^{j}$ denotes the $%
j $-th component of $v$. Combining with Eq. (\ref{me}), both sides of Eq. (%
\ref{pt1}) can be square-rooted and expanded as
\begin{equation}
d\tau =\sqrt{1+K}dt=\left( 1+\frac{K}{2}-\frac{K^{2}}{8}+\frac{K^{3}}{16}%
-\cdots \right) dt,\label{taut}
\end{equation}%
where
\begin{equation}
K\simeq \frac{1}{c^{2}}\left( -2\Phi (t,r)-v^{2}\right) +\frac{1}{c^{4}}%
\left[ 2\beta \Phi ^{2}(t,r)+\left( 4\gamma +1\right) \Phi
^{j}(t,r)v^{j}-2\gamma \Phi (t,r)v^{2}\right] . \label{K}
\end{equation}%
So we gain a brief relationship between the clock's proper time and
coordinate time in the LCRS that includes the lunar gravitational potential $%
\Phi _{m}(t,r)$, the tidal potential $\Phi _{t}(t,r)$ (mainly from the tidal
potential of the Earth $\Phi _{tE}(t,r)$ and the tidal potential of the Sun $%
\Phi _{ts}(t,r)$), the inertial potential $\Phi _{iner}(t,r)$ (caused by the
nongeodesic acceleration of the Moon center of mass), and the vector
potential $\Phi ^{j}(t,r)$,
\begin{equation}
\begin{aligned} \frac{d\tau }{dt}\simeq 1+\frac{-2\Phi
(t,r)-v^{2}}{2c^{2}}+\mathcal{O}\left(c^{-4}\right). \label{pt} \end{aligned}
\end{equation}

According to Eq. (\ref{pt}), it can be seen that the influence of the vector
potential on the proper time of clocks near the Moon is of the order of $%
c^{-4}$; consequently, the influence of the vector potential on these clocks
will be significantly smaller than $10^{-18}$. According to Eq. (\ref{taut}) and Eq. (\ref{K}), 
it can be seen that-at the post-Newtonian order $c^{-2}$ in Eq. (\ref{pt})-no cross-terms 
involving both the vector potential and the scalar potential appear. 
Consequently, the vector potential will not affect the LTR analyzed in our subsequent discussion. All calculations and analyses
in the next section will remain at the level of $10^{-18}$ according to the
measurement precision of the clock at present. In the LCRS, the orders of
magnitude of the various components of the vector potential have been
comprehensively estimated \cite{Bourgoin}; we will not discuss this in
further detail here. We will neglect the terms of order $c^{-4}$ and higher
in Eq. (\ref{pt}) in the subsequent calculations. The scalar inertial
potential $\Phi _{iner}(t,r)$ is among the $c^{-2}$ term. According to
Soffel et al \cite{soffel2003iau}, it can be expressed as
\begin{equation}
\Phi _{iner}(t,r)=Q(t)r,
\end{equation}%
where
\begin{equation}
Q(t)\propto J_{2}^{m}\sum_{A\neq m}\frac{GM_{A}a_{m}^{2}}{r_{Am}^{4}}
\end{equation}%
represents the non-geodesic acceleration of the Moon. $J_{2}^{m}=2.03\times
10^{-4}$ is the quadrupole moment of the Moon. The magnitudes of the scalar
inertial potentials contributed by the Earth and the Sun are $\Phi
_{iner-E}(t,r)/c^{2}=J_{2}^{m}\frac{GM_{E}a_{m}^{3}}{c^{2}r_{mE}^{4}}%
=2.16\times 10^{-22}$, and $\Phi _{iner-s}(t,r)/c^{2}=J_{2}^{m}\frac{%
GM_{s}a_{m}^{3}}{c^{2}r_{ms}^{4}}=3.15\times 10^{-27}$, respectively; since
both are less than $10^{-18}$, we have neglected them in Eq. (\ref{pt}).
Here, $r_{mE}$ is the distance from the Earth to the Moon, $r_{ms}$ is the
distance from the Sun to the Moon, and $G$ is the gravitational constant.

Note that the influence on the clock from the Moon's gravitational potential
can reach the level of $10^{-18}$. We require to make some further analyses
on this term given by
\begin{equation}
\Phi _{m}=\frac{GM_{m}}{r}\sum_{l=0}^{\infty }\left( \frac{a_{m}}{r}\right)
^{l}\sum_{q=0}^{l}\bar{P}_{l,q}(\sin \phi )\left[ \bar{C}_{l,q}\cos \left(
q\lambda \right) +\bar{S}_{l,q}\sin \left( q\lambda \right) \right] ,
\label{pe}
\end{equation}%
where $GM_{m}$ is the product of Moon's mass times the Newtonian
gravitational constant, $a_{m}=1738140m$ is Moon's equatorial radius, $%
r,\phi ,\lambda $ are the distance from a point on the Moon's surface to the
Moon's center of mass, the latitude of the point on the Moon's surface, and
the longitude, respectively. $\bar{C}_{l,q}$ and $\bar{S}_{l,q}$ are
normalized spherical harmonic coefficients of degree $l$ and order $q$, and $%
\bar{P}_{l,q}$ are the normalized associated Legendre functions of degree $l$
and order $q$. The detailed expression can be seen in Ref. \cite%
{konopliv2013jpl}. The order of magnitude of the lunar gravitational
potential is $3.14\times 10^{-11}$, and the order of magnitude of the
velocity term is $2.38\times 10^{-16}$ \cite{Bourgoin}. The velocity term is significantly larger than the precision we are concerned with, and since it cancels out when comparing clocks, we will not discuss the velocity term $\sim v^{2}/c^{2}$ in Eq. (\ref{pt}). It is noted
that the normalized spherical harmonic coefficients $\bar{C}_{l,q}$ and $%
\bar{S}_{l,q}$ in the Moon's gravitational potential (\ref{pe}) tend to
smaller values as the degree increases. This is partly due to the
characteristics of the Moon's gravitational field, but the main reason is
that the Legendre function tends to a larger value as the degree increases.
Consequently, the higher-order terms of the lunar gravitational potential
will decrease accordingly; specifically, the term $\Phi
_{m}|_{l=37,q=1}/c^{2}$ with degree $l=37$ and order $q=1$ has a magnitude
of $1\times 10^{-18}$ (where the longitude is taken as $\lambda =0^{\circ }$%
). Although the individual contributions of the higher-order terms of the
Moon's gravitational potential are only on the order of $1\times 10^{-18}$,
their cumulative effect, similar to that of Earth's gravitational potential,
can still be significant, even at the level of $1\times 10^{-16}$, which has
been studied before \cite{slava} and we will not consider it in Eq. (\ref{pt}%
) in our calculation.

Because the Moon is an elastic body, the tidal potential generated by the
non-uniformity of the Earth's and Sun's gravitational fields causes tidal
deformation. This tidal deformation is the response of the Moon to the
Earth's and Sun's gravitational fields, which can cause observable
perturbations in the satellite's orbital motion and clock rate \cite%
{qin2020tidal}. To consider this, we introduce tidal response into the LCRS.
In particular, this deformation produces a time-varying potential, which
contributes a correction to the metric tensor by changing Moon's
gravitational potential into
\begin{equation}
\Phi _{m}^{^{\prime }}=\Phi _{m}+\Phi _{tr},  \label{tr}
\end{equation}%
where $\Phi _{tr}=k_{2}^{\mathrm{m}}\Phi _{t}$ is the tidal response \cite%
{williams2015tides,williams2014lunar}, $k_{2}^{m}=0.02405$ is the lunar Love
number \cite{williams2014lunar}, which leads to a significant time-dependent
contribution to the lunar gravitational potential.

The tidal potential not only forms part of the metric (\ref{me}), but also
produces a tidal response that modifies the metric (\ref{me}). The
modification in Eq. (\ref{tr}) is permissible because, according to the $%
c^{-2}$ order terms in Eq. (\ref{pt}), the lunar gravitational potential $%
\Phi _{m}$, the tidal potential $\Phi _{t}$, and the inertial potential $%
\Phi _{iner}$ combine additively. Consequently, by drawing an analogy with
the modification of the GCRS metric regarding Earth's tidal response \cite%
{qin2020tidal,qin2023tidal}, we have applied a similar modification to the
LCRS. It should be noted that the lunar gravitational potential $\Phi _{m}$
corrected here is of order $c^{-2}$; consequently, the post-Newtonian order
of the LTR $\Phi _{tr}$ is also $c^{-2}$. For Moon, the measurement from the
gravimeter can be used to determine Love numbers, which reveal information
about the tidal response.

We will primarily explore the influence of the LTR caused by the Earth's
tidal field on the frequency of clocks located on the lunar surface. Before
studying the LTR, it is necessary to first get the tidal potential
experienced by the lunar surface. When the Moon moves in an approximately
circular orbit, the tide-raising potential generated by the Earth's
gravitational field at a point A at radius $r_{A}$ from the center of the
Moon can be expressed as \cite%
{murray1999solar,williams2015tides,williams2014lunar}
\begin{equation}
\Phi _{tE}=\frac{GM_{E}r_{A}^{2}}{2r_{mE}^{3}}\left( 3\cos ^{2}\psi
-1\right) +\frac{GM_{E}r_{A}^{3}}{2r_{mE}^{4}}\left( 5\cos ^{3}\psi -3\cos
\psi \right) +\cdots ,  \label{te}
\end{equation}%
where $M_{E}$ is the mass of the Earth, the angle $\psi $ is the angle
between the position vector $\mathbf{r_{A}}$ and the position vector $%
\mathbf{r_{mE}}$, and $r_{mE}$ is the distance from the center of the Moon
to the Earth's center. The magnitude of the influence of the second-order
term of the Earth's tidal potential is $2.35\times 10^{-16}$, which is far
greater than $10^{-18}$; therefore, we will not consider it. The influence
of the third-order term is $5.32\times 10^{-19}$, and its higher-order terms
are also less than $10^{-18}$. For tides raised by the Earth, analogous
higher-degree tides weaken by $2$ orders of magnitude per degree, so the
third-order and higher-order terms are neglected here. Therefore, the Earth
tidal potential term $\Phi _{tE}$ is neglected in Eq. (\ref{pt}).

The solar tidal potential and the Earth's tidal potential have similar forms
\cite{ashby2024relativistic,turyshev2025relativistic}
\begin{equation}
\Phi _{ts}=\frac{GM_{s}r_{A}^{2}}{2r_{ms}^{3}} \left( 3\cos ^{2}\psi_s
-1\right)+\frac{GM_{s}r_{A}^{3}}{2r_{ms}^{4}} \left( 5\cos ^{3}\psi_s
-3\cos\psi_s\right)+\cdots,  \label{ts}
\end{equation}
where $GM_{s}$ is the product of solar mass times the Newtonian
gravitational constant, $r_{ms}$ is the distance from the Moon to the Sun,
and the angle $\psi_s $ is the angle between the position vector $\mathbf{r_A%
}$ and the position vector $\mathbf{r_{ms}}$. The magnitude of the influence
of the second-order term of the solar tidal potential on a lunar clock is $%
6.64 \times 10^{-19}$, and its higher-order terms are also less than $%
10^{-18}$. Moreover, the Earth's orbital period around the Sun is much
longer than the Moon's orbital period around the Earth, and the Earth's
tidal potential on the Moon's surface is several times that of the Sun.
Therefore, we will neglect the solar tidal potential $\Phi _{ts}$ in Eq.(\ref%
{pt}).

Then, we continue to consider the case in which the Moon moves in an
elliptical orbit $r_{mE}=a\left( 1-e^{2}\right) /\left( 1+e\cos T\right) $
where $a$, $e$, and $T$ denote the semi-major axis, eccentricity, and true
anomaly of the Moon's orbit, respectively. When investigating the tidal
interactions experienced by clock A on the lunar surface as the Moon moves
along an elliptical orbit, it is necessary to consider the angle $\psi _{mA}$
between the position vector of clock A in the LCRS $\mathbf{r}_{A}$ and the
position vector of the Earth $\mathbf{r}_{mE}$. The position vector of clock
A is
\begin{equation}
\mathbf{r}_{A}=r_{A}\cos \phi _{A}\cos \lambda _{A}+r_{A}\cos \phi _{A}\sin
\lambda _{A}+r_{A}\sin \phi _{A},
\end{equation}%
where $r_{A}$ is the distance from the clock to the center of mass of the
Moon, $\phi _{A}$ is the latitude, and $\lambda _{A}$ is the longitude. The
position of the Earth relative to the Moon in the LCRS can be obtained by
rotating the position vector of the Moon. First, select a position vector
\begin{equation}
\mathbf{u}=\boldsymbol{x}_{m}\cos \left( T+\omega \right) +\boldsymbol{y}%
_{m}\sin \left( T+\omega \right) ,
\end{equation}%
lying within the lunar orbital plane, where $T$ denotes the true anomaly and
$\omega $ represents the argument of perigee. In Fig. 1, the direction of
the $x_{m}$-axis corresponds to the direction of $\mathbf{u}$. Rotate by an
angle $\epsilon $ around the $x_{m}$-axis to shift the vector from the lunar
orbital plane to the ecliptic plane. Then, rotate by an angle $L-\Omega $
around the $z_{m}$-axis to align the vector with the direction of the Earth (%
$x_{E}$-axis). Here, $\epsilon $ represents the orbital inclination, $L$ is
the mean longitude, and $\Omega $ is the longitude of the ascending node.
These two rotations can be realized through matrices $R_{x_{m}}\left(
\epsilon \right) $ and $R_{z_{m}}\left( L-\Omega \right) $, and the position
vector $\mathbf{r}_{mE}$ can be derived via $\mathbf{r}_{mE}=r_{mE}R_{z_{m}}%
\left( L-\Omega \right) R_{x_{m}}\left( \epsilon \right) \mathbf{u}$ as
\begin{equation}
\begin{aligned} \mathbf{r}_{mE}=& r_{mE}\left[\cos \epsilon\sin
\left(L-\Omega\right) \sin \left(T+\omega\right)+\cos \left(L-\Omega\right)
\cos \left(T+\omega\right)\right] -r_{mE} \sin \epsilon \sin
\left(T+\omega\right)\\ & +r_{mE}\left[\cos \epsilon \cos
\left(L-\Omega\right) \sin \left(T+\omega\right)-\sin \left(L-\Omega\right)
\cos \left(T+\omega\right)\right] , \end{aligned}
\end{equation}%
where
\begin{equation}
R_{x_{m}}\left( \epsilon \right) =\left(
\begin{array}{ccc}
1 & 0 & 0 \\
0 & \cos \epsilon & \sin \epsilon \\
0 & -\sin \epsilon & \cos \epsilon%
\end{array}%
\right), \text{and} \quad R_{z_{m}}\left( L-\Omega \right) =\left(
\begin{array}{ccc}
\cos \left( L-\Omega \right) & \sin \left( L-\Omega \right) & 0 \\
-\sin \left( L-\Omega \right) & \cos \left( L-\Omega \right) & 0 \\
0 & 0 & 1%
\end{array}%
\right) .
\end{equation}

\begin{figure}[ptb]
\centering
{\includegraphics[width=0.6\textwidth]{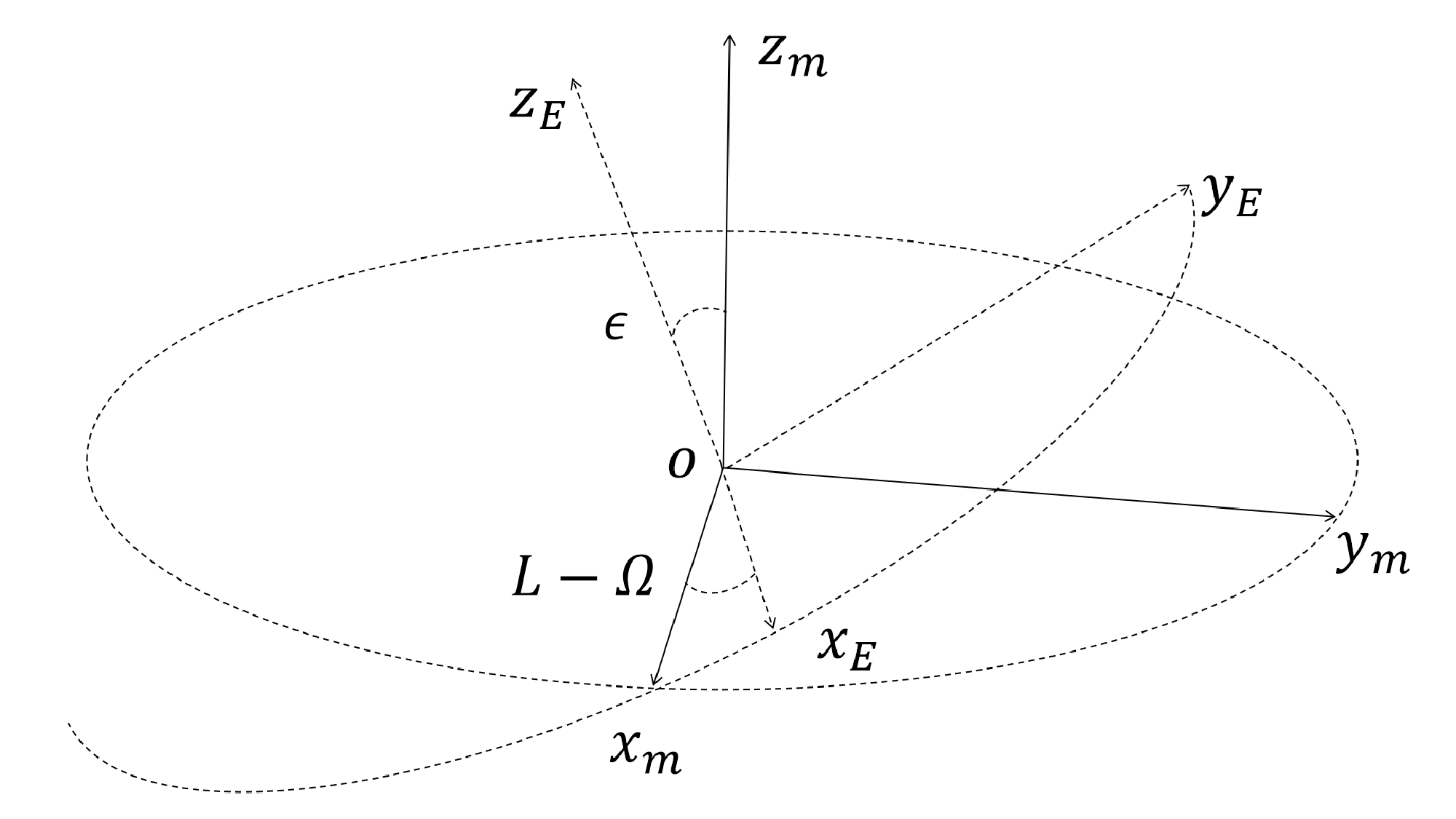}}
\caption{A schematic diagram illustrating the direction of Earth relative to
the LCRS. The subscript $E$ represents Earth, and the subscript $m$
represents the Moon.}
\label{Fig1}
\end{figure}

Then we can gain the angle $\psi $ as follows,
\begin{equation}
\cos \psi =\frac{\mathbf{r}_{A}\mathbf{r}_{mE}}{r_{A}r_{mE}}=C_{A}\cos
2\left( T+\omega \right) +D_{A}\sin 2\left( T+\omega \right) ,  \label{ag}
\end{equation}%
where
\begin{equation}
C_{A}=\cos \phi _{A}\cos \left( L-\Omega +\phi _{A}\right), \quad \text{and}
\quad D_{A}=\cos \phi _{A}\cos \epsilon \sin \left( L-\Omega +\phi
_{A}\right) -\sin \phi _{A}\sin \epsilon .
\end{equation}%
By combining Eq. (\ref{te}) and (\ref{ag}) with the Kepler expansions \cite%
{murray1999solar}
\begin{equation}
\frac{1}{r_{mE}^{3}}=\frac{1}{a^{3}}\left[ 1+3e\cos M+3e^{2}\left( 2\cos
^{2}M-\sin ^{2}M\right) +\mathcal{O}\left( e^{-3}\right) \right]  \label{r}
\end{equation}%
and
\begin{equation}
T=M+2e\sin M+\frac{5}{4}e^{2}\sin 2M+\mathcal{O}\left( e^{-3}\right) ,
\label{T}
\end{equation}%
we can derive an expression for the Earth's tidal potential acting on Clock
A, located on the Moon as it follows an elliptical orbit. Here, $M$ is the
mean anomaly of the lunar orbit. Presented here is the second-order term of
the tidal potential,
\begin{equation}
\begin{split}
\Phi _{tE2}& =\frac{GM_{E}r_{A}^{2}}{2a^{3}}\left[ \frac{3}{2}\left(
C_{A}^{2}+D_{A}^{2}\right) +\frac{3}{2}\left( C_{A}^{2}-D_{A}^{2}\right)
\cos \left( 2M+2\omega \right) -1+3C_{A}D_{A}\sin \left( 2M+2\omega \right) %
\right] \\
& \times \left( 1+3e\cos M\right) +\frac{GM_{E}r_{A}^{2}}{a^{3}}\left[
6eC_{A}D_{A}\sin M\cos \left( 2M+2\omega \right) -3e\sin M\sin \left(
2M+2\omega \right) \right. \\
& \left. \times \left( C_{A}^{2}-D_{A}^{2}\right) \right] .
\end{split}
\label{te2}
\end{equation}%
The higher-order terms of eccentricity are ignored here and later in the
paper, because the eccentricity of the Moon's orbit $e=0.0549006$ is small.

From the angular relations (\ref{ag}), the Keplerian expansions (\ref{r})
and (\ref{T}), and Eqs. (\ref{tr}), (\ref{te}), and (\ref{ts}), we can
estimate the magnitude of the tidal response induced on the Moon by the
tidal potential. The contributions of all terms of the solar tidal potential
are less than $10^{-18}$; therefore, according to Eq. (\ref{tr}), the
contributions of the tidal response induced by the solar tidal potential is
significantly less than $10^{-18}$. Consequently, we have neglected this
component of the tidal response in Eq. (\ref{pt}), and $\Phi _{tr}$
henceforth denotes solely the tidal response induced by the Earth's tidal
potential. The contribution of the tidal response induced by the
second-order term of the Earth's tidal potential is $2.81\times 10^{-18}$;
the contribution of the higher-order terms of the Earth's tidal potential is
smaller than $10^{-18}$, implying that the higher-order terms of the induced
tidal response are significantly smaller than $10^{-18}$. Therefore, we
consider only the tidal response induced by the second-order term of the
Earth's tidal potential Eq. (\ref{te2}) and neglect the higher-order terms
of the tidal response in Eq. (\ref{pt}).

\begin{table}[tbp]
\caption{Potentials and order of magnitude.}
\label{tab:const2}\centering
\renewcommand{\arraystretch}{1.2}
\begin{tabular}{p{11.6cm}p{4.4cm}}
\hline\hline
Potential & Order of magnitude \\ \hline
Scalar inertial potentials contributed by the Earth & $\frac{\Phi_{iner-E}}{%
c^{2}}=2.16\times10^{-22}$ \\
Scalar inertial potentials contributed by the Sun & $\frac{\Phi_{iner-s}}{%
c^{2}}=3.15\times10^{-27}$ \\
Lunar gravitational potential & $\frac{\Phi_{m}}{c^{2}}=3.14\times 10^{-11}$
\\
Velocity term & $\frac{v^2}{c^2}=2.38\times10^{-16}$ \\
Second-order term of the Earth's tidal potential & $\frac{\Phi_{tE2}}{c^{2}}%
=1.17\times10^{-16}$ \\
Third-order term of the Earth's tidal potential & $\frac{\Phi_{tE3}}{c^{2}}%
=5.32\times10^{-19}$ \\
Second-order term of the Sun's tidal potential & $\frac{\Phi_{ts2}}{c^{2}}%
=6.64\times10^{-19}$ \\
LTR to the second-order term of the Earth's tidal potential & $\frac{%
\Phi_{tr-tE2}}{c^{2}}=2.81\times10^{-18}$ \\
LTR to the third-order term of the Earth's tidal potential & $\frac{%
\Phi_{tr-tE3}}{c^{2}}=1.28\times10^{-20}$ \\
LTR to the second-order term of the Sun's tidal potential & $\frac{%
\Phi_{tr-ts2}}{c^{2}}=1.60\times10^{-20}$ \\ \hline
\end{tabular}%
\end{table}

Finally, we outline the hierarchy of approximations employed: under the
weak-field general relativity, we neglect contributions from terms of order $%
c^{-4}$ and higher; furthermore, within the Newtonian expansion of the
gravitational potential, we disregard terms whose contributions fall below
the order of $10^{-18}$. Moving a step further, we model the Moon's
trajectory as an ellipse, performing approximate expansions for both the
true anomaly and the Earth-Moon distance. When considering tidal response,
we have neglected tidal dissipation. This is because tidal dissipation is a
factor that must be taken into account when examining long-term evolutionary
processes \cite{williams2015tides}-such as those leading to changes in the
Moon's orbital eccentricity-as well as issues related to thermal evolution
\cite{murray1999solar}. However, in the context of this paper, which focuses
on short-term clock comparison problems, the effects of tidal dissipation
are not relevant. Since we neglected tidal dissipation, the paper does not
include tidal lag effects. Based on the current precision of atomic clocks,
all calculations and analyses in the next section will be maintained at a
precision level of $10^{-18}$. Therefore, it is appropriate to consider the
influence of the LTR caused by the Earth's tidal potential on clocks near
the Moon. See Table II for the numerical estimation of all potential terms
involved in this paper, and in the following, we will focus on the influence
of LTR caused by the second-order term of the Earth's tidal potential on the
clock's FFS.

\section{Frequency shift between two clocks}

In this section, we consider the FFS between two clocks at different
locations on the lunar surface. The FFS between two clocks on the lunar
surface is affected by the Moon's own tidal response. When we consider the
clocks A and B at different locations, we can calculate the FFS between them
by using Eqs. (\ref{pt}) and (\ref{tr}) as
\begin{equation}
\frac{\triangle f}{f_{B}}=\frac{f_{A}-f_{B}}{f_{B}}=\frac{d\tau _{B}}{d\tau
_{A}}-1=\frac{1+P_{B}}{1+P_{A}}-1,  \label{fff}
\end{equation}%
where $f_{A}=1/d\tau _{A}$, $f_{B}=1/d\tau _{B}$, $P_{A}=-\Phi _{trA}/c^{2}$
and $P_{B}=-\Phi _{trB}/c^{2}$. We can expand Eq. (\ref{fff}) to get a new
form of FFS as
\begin{eqnarray}
\frac{\triangle f}{f_{B}} &=&\left[ 1+P_{B}-\left( 1+P_{B}\right)
P_{A}+\left( 1+P_{B}\right) P_{A}^{2}+\cdots \right] -1  \notag \\
&=&P_{B}-P_{A}+\mathcal{O}\left( c^{-4}\right) \simeq \frac{1}{c^{2}}\left(
\Phi _{trA}-\Phi _{trB}\right) .  \label{FFF}
\end{eqnarray}

Due to the difference in geographical latitude and longitude between clocks
A and B, the tidal response between the two clocks is different. Since the
above quantity is related to the specific location of the clock, there will
be an obvious position-dependent effect when we compare clock A and clock B
at two different locations on the lunar surface. We use the longitude $%
\lambda _{A}$ and latitude $\phi _{A}$ of the lunar surface clock to
determine the angle $\psi _{mA}$ between the position vector of the clock on
the lunar surface $\mathbf{r}_{A}$ and the position vector of the Earth $%
\mathbf{r}_{mE}$. Compared to previous methods \cite%
{qin2020tidal,qin2023tidal}, our method can analyze the change in FFS caused
by the longitude difference between the two clocks. Furthermore, using the
Kepler expansions method, we can also analyze the FFS between two clocks on
the lunar surface when the Moon is orbiting in an elliptical orbit.

Based on the estimates of each term in Eq. (\ref{pt}) above, the magnitude
of the LTR-caused FFS between the two clocks is $10^{-18}$. This magnitude
is comparable to the instability and uncertainty of the FFS achieved by
currently state-of-the-art clocks \cite%
{bloom2014optical,nicholson2015systematic,huntemann2016single,nakamura2020coherent,boulder2021frequency,marshall2025high}%
. Therefore, the FFS between the two clocks on the lunar surface caused by
the LTR can be measured. Using Eqs. (\ref{tr}), (\ref{te2}), and (\ref{FFF}%
), the LTR-caused FFS between clock A and clock B is
\begin{equation}
\begin{split}
\frac{\triangle f}{f_{B}}|_{trAB}& =\frac{k_{2}^{\mathrm{m}}GM_{E}r_{A}^{2}}{%
2c^{2}a^{3}}\left( 1+3e\cos M\right) \left[ \frac{3}{2}\left(
C_{A}^{2}+D_{A}^{2}\right) +\frac{3}{2}\left( C_{A}^{2}-D_{A}^{2}\right)
\cos \left( 2M+2\omega \right) -1\right. \\
& \left. +3C_{A}D_{A}\sin \left( 2M+2\omega \right) \right] +\frac{k_{2}^{%
\mathrm{m}}GM_{E}r_{A}^{2}}{c^{2}a^{3}}\left[ 6eC_{A}D_{A}\sin M\cos \left(
2M+2\omega \right) -3e\sin M\right. \\
& \left. \times \left( C_{A}^{2}-D_{A}^{2}\right) \sin \left( 2M+2\omega
\right) \right] -\frac{k_{2}^{\mathrm{m}}GM_{E}r_{B}^{2}}{2c^{2}a^{3}}\left(
1+3e\cos M\right) \left[ \frac{3}{2}\left( C_{B}^{2}+D_{B}^{2}\right) +\frac{%
3}{2}\right. \\
& \left. \times \left( C_{B}^{2}-D_{B}^{2}\right) \cos \left( 2M+2\omega
\right) -1+3C_{B}D_{B}\sin \left( 2M+2\omega \right) \right] -\frac{k_{2}^{%
\mathrm{m}}GM_{E}r_{B}^{2}}{c^{2}a^{3}}\left[ 6eC_{B}\right. \\
& \left. \times D_{B}\sin M\cos \left( 2M+2\omega \right) -3e\left(
C_{B}^{2}-D_{B}^{2}\right) \sin M\sin \left( 2M+2\omega \right) \right] .
\end{split}
\label{trab}
\end{equation}%
We consider the Moon's tidal response as elastic without dissipation, so the
Love number $k_{2}^{\mathrm{m}}=0.02405$ is sufficient. Tidal dissipation
from the lunar interior due to Earth-induced tides on the Moon can both heat
the Moon and circularize its orbit \cite{murray1999solar}. Unlike the case
discussed in this paper, the Love number for tidal dissipation is a complex
number \cite{williams2014lunar}.

\begin{figure}[ptb]
\centering
\begin{subfigure}[h]{0.48\textwidth}
		{\includegraphics[width=\textwidth]{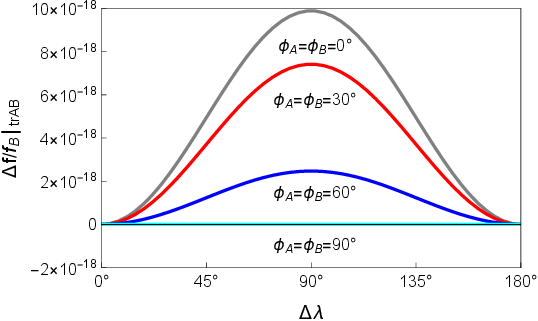}}
		\caption{}
		\label{2a}
	\end{subfigure}
\begin{subfigure}[h]{0.49\textwidth}
		{\includegraphics[width=\textwidth]{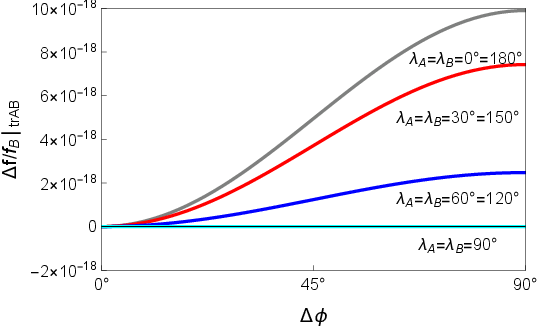}}
		\caption{}
		\label{2b}
	\end{subfigure}
\caption{The LTR-caused FFS between clocks A and B, which (a) have the same
latitude $\protect\phi_A=\protect\phi_B=0^\circ, 30^\circ, 60^\circ,
90^\circ $ but different longitudes, and (b) have the same longitude $%
\protect\lambda_A=\protect\lambda_B=0^\circ,30^\circ,60^\circ,
90^\circ,120^\circ,150^\circ,180^\circ$ but different latitudes. Other
parameters are given in SI units: $M_E=5.977\times 10^{24}$, $%
G=6.674\times10^{-11}$, $c=3\times 10^8$, $M=0^\circ$, $L-\Omega=0^\circ$, $%
\protect\epsilon=6.87^\circ$, $\protect\omega=0^\circ$, $k^m_2=0.02405$, $%
r_A=r_B=1.7371 \times 10^6$, $e=0.0549006$, and $a=3.84399\times 10^8$.}
\label{Fig2}
\end{figure}

In Fig. 2, we analyze the FFS between clocks at two different locations on
the lunar surface caused by the Moon's response to Earth's tides. For the
convenience of comparison, we consider that the longitude $\lambda _{A,B}$
range for the clock on the lunar surface is from $0^{\circ }$ to $180^{\circ
}$, and the latitude $\phi _{A,B}$ range is from $0^{\circ }$ to $90^{\circ
} $. For the models under consideration, we make the following assumptions:
the Moon is spherical, there is no physical libration, the lunar rotation
axis is aligned with the tidal axis, no precession of the line of apsides or
nodes occurs, and the Earth is fixed in the lunar coordinate frame. In Fig.
2a, clocks A and B are located at the same latitude, with clock A placed at
the longitude of zero degrees, compared to clock B at a different longitude.
The FFS is symmetrical around $90$ degrees longitude, and gradually
decreases as clock B moves away from $90$ degrees longitude. When we fix
clock B at a longitude, that is, when $\triangle \lambda $ is fixed, the
difference in tidal response between the two clocks decreases as their
latitude increases, resulting in a decrease in the FFS. When the latitudes
of clocks A and B are both increased to $90$ degrees, they are both located
at the lunar north or south pole, and there is no FFS. It is clearer to see
these by simplifying Eq. (\ref{trab}),
\begin{equation}
\frac{\triangle f}{f_{B}}|_{fig.2a}=9.89\times 10^{-18}\cos ^{2}\left( \phi
_{A/B}\right) \left[ 1-\cos ^{2}\left( \triangle \lambda \right) \right] .
\label{eq.2a}
\end{equation}%
This is the result of the function $\cos ^{2}\left( \phi _{A/B}\right) $
modulation on the amplitude term of the LTR-caused FFS.

\begin{figure}[ptb]
\centering
{\includegraphics[width=0.76\textwidth]{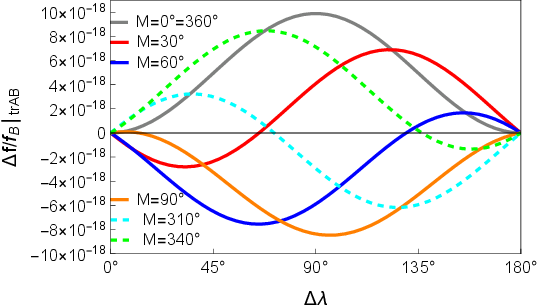}}
\caption{The LTR-caused FFS between clocks A and B with different lunar mean
anomaly $M=0^\circ,30^\circ,60^\circ,90^\circ, 310^\circ,340^\circ
,360^\circ $, where the same latitude $\protect\phi_A=\protect\phi_B=0^\circ$
but different longitudes are chosen. Other parameters are given in SI units:
$M_E=5.977\times 10^{24}$, $G=6.674\times10^{-11}$, $c=3\times 10^8$, $%
k^m_2=0.02405$, $L-\Omega=0^\circ$, $\protect\epsilon=6.87^\circ$, $\protect%
\omega=0^\circ$, $r_A=r_B=1.7371\times 10^6$, $e=0.0549006$, and $%
a=3.84399\times 10^8$.}
\label{Fig3}
\end{figure}

When considering two clocks A and B located at the same lunar longitude,
with clock A at $0$ degrees latitude and clock B at a different latitude,
the FFS increases as the latitude of clock B increases, as presented in Fig.
2b, due to the increasing difference in the tidal response felt by the two
clocks A and B. When clock B is also at a fixed latitude, that is, there is
a fixed latitude difference $\triangle \phi $ between clocks A and B, the
FFS decreases as the longitude of the two clocks increases from $0$ to $90$
degrees, or decreases from $180$ to $90$ degrees. This is because the LTR
felt by the two clocks becomes increasingly similar. It can be seen that the
function graph of the LTR-caused FFS in the figure exactly matches the
equation%
\begin{equation}
\frac{\triangle f}{f_{B}}|_{fig.2b}=9.89\times 10^{-18}\cos ^{2}\left(
\lambda _{A/B}\right) \left[ 1-\cos ^{2}\left( \triangle \phi \right) \right]
,  \label{eq.2b}
\end{equation}%
which is obtained by simplifying Eq. (\ref{trab}). The line $\lambda
_{A}=\lambda _{B}=90^{\circ }$ in Fig. 2b is a special case. Since the mean
anomaly of the Moon is fixed at $M=0^{\circ }$, when the longitude of the
two clocks is 90 degrees, the LTR at clocks A and B is $0$, and there is no
LTR-caused FFS.

\begin{figure}[ptb]
\centering
{\includegraphics[width=0.76\textwidth]{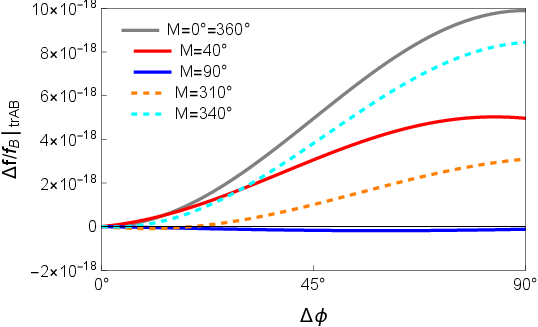}}
\caption{The LTR-caused FFS between clocks A and B with different lunar mean
anomaly $M=0^\circ,40^\circ,90^\circ, 310^\circ,340^\circ ,360^\circ$, where
different latitudes but the same longitude $\protect\lambda_A=\protect\lambda%
_B=0^\circ$ are chosen. Other parameters are kept consistent with those in
Fig. 3.}
\label{Fig4}
\end{figure}

In Fig. 2, we consider the FFS due to the tidal response of the Moon between
two clocks at different locations on the lunar surface. With the help of the
Kepler expansions method, we can couple the orbital elements of the Moon's
motion into the LTR, so that we can analyze the effect of the Moon's motion
on the LTR-caused FFS. In Fig. 3 and Fig. 4, the mean anomaly of the Moon is
no longer zero. The mean anomaly of the Moon will change from $0^{\circ }$
to $360^{\circ }$ during one cycle of its elliptical orbit, so here and
later we only consider the range of lunar mean anomaly changes from $%
0^{\circ }$ to $360^{\circ }$. However, the change in the mean anomaly of
the lunar orbit has different effects on the LTR-caused FFS. In Fig. 3, it
is seen that the phase and amplitude of the FFS between two clocks with only
a longitude difference change periodically with the motion of the Moon. A
more intuitive reason can be seen by the analyzed expression%
\begin{equation}
\frac{\triangle f}{f_{B}}|_{fig.3}\simeq 4.21\times 10^{-18}\left[ \cos
2M-\cos 2\left( M-\triangle \lambda \right) \right] ,
\end{equation}%
through simplifying Eq. (\ref{trab}). In this case, changing the mean
anomaly of the Moon will simultaneously change the phase and amplitude of
the FFS function. However, when the Moon moves, the FFS between two clocks
with only a latitude difference changes only in amplitude, as seen in Fig.
4. The reason for this change can be seen clearly by simplifying Eq. (\ref%
{trab}) into%
\begin{equation}
\frac{\triangle f}{f_{B}}|_{fig.4}\simeq 2.15\times 10^{-18}\cos
2M+2.09\times 10^{-18}\left( 1-\cos 2\triangle \phi \right) .
\end{equation}%
Changing the mean anomaly of the Moon only changes the amplitude term $\cos
\left( 2M\right) $ of the LTR-related FFS function.

\section{Clock differences caused by LTR}

As humanity seeks to explore the solar system and study distant planets like
the Moon and Mars, there is a growing need to estimate the clock rates on
these celestial bodies and compare them to the standard clock rate on Earth.
Recently, as part of establishing a framework for a lunar time standard,
Ashby and Patla calculated the rate at which clocks would run on the Moon
and proposed a method for synchronizing Earth clocks with lunar clocks \cite%
{ashby2024relativistic}. Establishing a unified, common lunar time standard
is crucial for achieving technological synchronization of lunar-based
infrastructure, promoting the development of a lunar economy, and ensuring
the smooth conduct of scientific research activities on the lunar surface.
Using this method, they determined that time on the Moon runs 56.02 $\mu $s
faster than on Earth every day. Providing an estimate of relativistic
effects on clocks on the lunar surface will allow us to compare their
accuracy when actual clocks are placed on the Moon.

Like Ashby and Patla, Kopeikin and Kaplan also performed similar
calculations \cite{kopeikin2024lunar}, and these researchers found that time
on the Moon runs, on average, 56 $\mu $s faster per day than on Earth. In
addition to considering the effects of the Earth's and Moon's gravitational
fields, Kopeikin and Kaplan further considered the effects of solar and
Earth tides on the rate of clocks on the Moon's surface, which cause the
secular drift of clocks of $8.7\times 10^{-8}$ $\mu $s/day and $1.5\times
10^{-5}$ $\mu $s/day, respectively.

Here, we can estimate the effect of LTR on the clock rate at the lunar
surface using Eq. (\ref{pt}) and Eq. (\ref{tr}). The secular drift of clocks
caused by the LTR is $8.6\times 10^{-7}$ $\mu $s/day. The LTR causes the
secular drift of the clock on the Moon's surface that is smaller than that
caused by Earth's tides but larger than that caused by the Sun's tides.
Therefore, LTR is an important factor to consider when establishing a lunar
time standard. In this section, we will consider the effect of LTR on clock comparisons near the Moon. According to Eq. (\ref{fff}),
we have the following expression
\begin{equation}
\frac{d\tau _{B}-d\tau _{A}}{d\tau _{A}}=\frac{\triangle f}{f_{B}}.
\label{dt}
\end{equation}%
Multiplying the above result by $H=10^{6}\times 86,400$ $\mu $s gives the
daily time difference between two clocks \cite%
{kopeikin2024lunar,ashby2024relativistic} at different locations on the
lunar surface. Furthermore, by means of Eq. (\ref{dt}), we can express the
time difference between two clocks in the vicinity of the Moon caused by the
LTR
\begin{equation}
\triangle \tau =\int_{0}^{H}\frac{d\tau _{B}-d\tau _{A}}{d\tau _{A}}dt=\frac{%
\triangle f}{f_{B}}H.  \label{dayt}
\end{equation}%
During the Moon's motion, only the mean anomaly changes, which does not
affect our estimation of time difference because the mean anomaly will take
a fixed value when estimating time. In the previous section, we considered
the FFS between two clocks at different longitudes and latitudes on the
lunar surface due to the LTR. Here, we can estimate the range of clock
differences between two clocks at different locations on the lunar surface
caused by the LTR during the Moon's motion to be $0\sim 8.64\times 10^{-7}$ $%
\mu $s/day.

\begin{figure}[ptb]
\centering
{\includegraphics[width=0.76\textwidth]{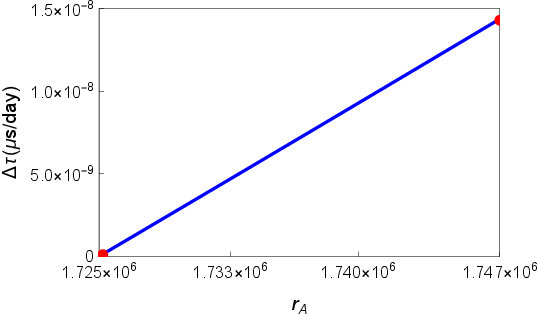}}
\caption{The time difference between a clock at the lowest point on the
lunar surface $(r_B=1725812 m)$ and a clock at a different distance from the
lunar center, caused by the LTR. The two red dots represent the highest $%
(r_A=1747729 m)$ and lowest $(r_A=1725812 m)$ points on the lunar surface,
respectively. Other parameters are given in SI units: $M_E=5.977\times
10^{24}$, $G=6.674\times10^{-11}$, $c=3\times 10^8$, $k^m_2=0.02405$, $%
e=0.0549006$, $\protect\phi _{A}=\protect\phi _{B}=\protect\lambda _{A}=%
\protect\lambda _{B}=M=0^\circ$, $L-\Omega=0^\circ$, $\protect\epsilon%
=6.87^\circ$, $\protect\omega=0^\circ$, and $a=3.84399\times 10^8$.}
\label{Fig5}
\end{figure}

Furthermore, we can estimate the clock difference caused by the LTR at
different distances from the Moon's center of mass. In Fig. 5, we consider
the distance from the lowest point on the lunar surface \cite{Araki,Li} to
the Moon's center of mass as the lower limit and the distance from the
highest point on the lunar surface \cite{Li} to the Moon's center of mass as
the upper limit. We consider clock B to be fixed at the lower limit point,
and then observe the clock difference between clock A and clock B as clock A
gradually moves from the lower limit point to the upper limit point, due to
the LTR. The time difference between two clocks, caused by the LTR,
increases linearly with the difference in their distances from the Moon's
center of mass. For comparison, we set the longitude, latitude, and mean
anomaly of the Moon for both clocks to $0^{\circ }$ using Eq. (\ref{trab}).
If the longitude, latitude, and distance from the Moon's center of mass for
the two clocks on the lunar surface are accurately known, the time
difference between the two clocks caused by the LTR can be estimated using
Eq. (\ref{trab}) and Eq. (\ref{dayt}). When clock A is actually at the
highest point on the lunar surface $(158.656^{\circ }W,5.441^{\circ }N)$ and
clock B is actually at the lowest point on the lunar surface $%
(172.413^{\circ }W,70.368^{\circ }S)$, the time difference between the two
clocks caused by lunar tidal effects is $6.43148\times 10^{-7}$ $\mu $s/day.

\begin{figure}[ptb]
\centering
{\includegraphics[width=0.76\textwidth]{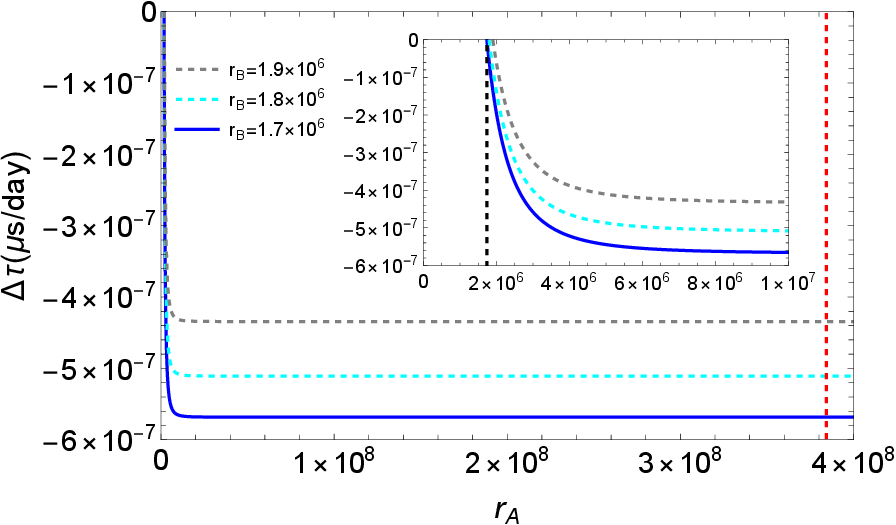}}
\caption{The solid line represents the time difference, induced by the LTR
effect, between a clock located on the lunar surface ($r_B=1.7371%
\times10^{6} m$) and a clock on a satellite gradually moving away from the
lunar surface. The dashed line represents the time difference between a
clock on a satellite located at a distance of $(r_B=1.8\times10^{6} m)$ or $%
(r_B=1.9\times10^{6} m)$ from the center of the Moon, and a clock on a
satellite gradually moving away from the lunar surface; this time difference
is also caused by the LTR effect. In the upper right corner, we show a
magnified image of the region where $r_A$ ranges from $0m$ to $%
1\times10^{7}m $. The black dashed line represents the distance from the
lunar surface to the Moon's center of mass, and the red dashed line
represents the distance from the Earth to the Moon's center of mass. Other
parameters are kept consistent with those in Fig. 5.}
\label{Fig6}
\end{figure}

The above analysis regarding the influence of LTR on clocks on the lunar
surface is important for achieving higher-precision time synchronization and
positioning for missions such as Chang'e 4 \cite{Wu} and Blue Ghost \cite%
{Sciacca} on the lunar surface. Performing higher-precision time corrections
for clocks on the lunar surface and on lunar satellites is equally
important. Lunar satellites can not only provide positioning, navigation,
and time synchronization services for the Moon, but also establish relay
communication between the Moon and the Earth. American researchers proposed
many schemes for lunar relay communication satellites a long time ago \cite%
{Farquhar,Burns,Mimoun}, and Chinese researchers also proposed schemes for
lunar relay communication satellites \cite{Tang,Wu2} and went even further
by successfully launching the Queqiao relay satellite \cite{zhang3}.
Therefore, we expanded the scope of our analysis to derive the LTR
experienced by a lunar satellite \cite{williams2015tides} by extending Eq. (%
\ref{te2})
\begin{equation}
\begin{split}
\Phi _{tr}^{\prime }& =\frac{k_{2}^{\mathrm{m}}GM_{E}r_{m}^{5}}{%
2a^{3}r_{A}^{3}}\left( 1+3e\cos M\right) \left[ \frac{3}{2}\left(
C_{A}^{2}+D_{A}^{2}\right) +\frac{3}{2}\left( C_{A}^{2}-D_{A}^{2}\right)
\cos \left( 2M+2\omega \right) -1+3C_{A}\right. \\
& \left. \times D_{A}\sin \left( 2M+2\omega \right) \right] +\frac{k_{2}^{%
\mathrm{m}}GM_{E}r_{m}^{5}}{a^{3}r_{A}^{3}}\left[ 6eC_{A}D_{A}\sin M\cos
\left( 2M+2\omega \right) -3e\sin \left( 2M+2\omega \right) \right. \\
& \left. \times \left( C_{A}^{2}-D_{A}^{2}\right) \sin M\right] .
\end{split}
\label{tr1}
\end{equation}%
where $r_{m}=1737100m$ is the average radius of the Moon. In this way, we
can obtain the time difference between clock B on the lunar surface and
clock A on the lunar satellite using equations Eq. (\ref{dayt}) and Eq. (\ref%
{tr1}), which is caused by the LTR
\begin{equation}
\begin{split}
\triangle \tau |_{ms}& =\frac{Hk_{2}^{\mathrm{m}}GM_{E}r_{m}^{5}}{%
2c^{2}a^{3}r_{A}^{3}}\left( 1+3e\cos M\right) \left[ \frac{3}{2}\left(
C_{A}^{2}+D_{A}^{2}\right) +\frac{3}{2}\left( C_{A}^{2}-D_{A}^{2}\right)
\cos \left( 2M+2\omega \right) -1\right. \\
& \left. +3C_{A}D_{A}\sin \left( 2M+2\omega \right) \right] +\frac{Hk_{2}^{%
\mathrm{m}}GM_{E}r_{m}^{5}}{c^{2}a^{3}r_{A}^{3}}\left[ 6eC_{A}D_{A}\sin
M\cos \left( 2M+2\omega \right) -3e\sin M\right. \\
& \left. \times \left( C_{A}^{2}-D_{A}^{2}\right) \sin \left( 2M+2\omega
\right) \right] -\frac{Hk_{2}^{\mathrm{m}}GM_{E}r_{B}^{2}}{2c^{2}a^{3}}%
\left( 1+3e\cos M\right) \left[ \frac{3}{2}\left( C_{A}^{2}+D_{A}^{2}\right)
+\frac{3}{2}\right. \\
& \left. \times \left( C_{A}^{2}-D_{A}^{2}\right) \cos \left( 2M+2\omega
\right) -1+3C_{A}D_{A}\sin \left( 2M+2\omega \right) \right] -\frac{Hk_{2}^{%
\mathrm{m}}GM_{E}r_{B}^{2}}{c^{2}a^{3}}\left[ 6eC_{A}\right. \\
& \left. \times D_{A}\sin M\cos \left( 2M+2\omega \right) -3e\sin M\sin
\left( 2M+2\omega \right) \left( C_{A}^{2}-D_{A}^{2}\right) \right] .
\end{split}
\label{trms}
\end{equation}%
Furthermore, we can obtain the time difference between clock B and clock A
both on the lunar satellite, which is caused by the LTR
\begin{equation}
\begin{split}
\triangle \tau |_{ss}& =\frac{Hk_{2}^{\mathrm{m}}GM_{E}r_{m}^{5}}{%
2c^{2}a^{3}r_{A}^{3}}\left( 1+3e\cos M\right) \left[ \frac{3}{2}\left(
C_{A}^{2}+D_{A}^{2}\right) +\frac{3}{2}\left( C_{A}^{2}-D_{A}^{2}\right)
\cos \left( 2M+2\omega \right) -1\right. \\
& \left. +3C_{A}D_{A}\sin \left( 2M+2\omega \right) \right] +\frac{Hk_{2}^{%
\mathrm{m}}GM_{E}r_{m}^{5}}{c^{2}a^{3}r_{A}^{3}}\left[ 6eC_{A}D_{A}\sin
M\cos \left( 2M+2\omega \right) -3e\sin M\right. \\
& \left. \times \left( C_{A}^{2}-D_{A}^{2}\right) \sin \left( 2M+2\omega
\right) \right] -\frac{Hk_{2}^{\mathrm{m}}GM_{E}r_{m}^{5}}{%
2c^{2}a^{3}r_{B}^{3}}\left( 1+3e\cos M\right) \left[ \frac{3}{2}\left(
C_{A}^{2}+D_{A}^{2}\right) +\frac{3}{2}\right. \\
& \left. \times \left( C_{A}^{2}-D_{A}^{2}\right) \cos \left( 2M+2\omega
\right) -1+3C_{A}D_{A}\sin \left( 2M+2\omega \right) \right] -\frac{Hk_{2}^{%
\mathrm{m}}GM_{E}r_{m}^{5}}{c^{2}a^{3}r_{B}^{3}}\left[ 6eC_{A}\right. \\
& \left. \times D_{A}\sin M\cos \left( 2M+2\omega \right) -3e\sin M\sin
\left( 2M+2\omega \right) \left( C_{A}^{2}-D_{A}^{2}\right) \right] .
\end{split}
\label{trss}
\end{equation}

From the blue solid line in Fig. 6, we can see that the time difference
between the clock B on the lunar surface and the clock A on the satellite,
caused by the LTR, gradually increases (that means the absolute values) as
the satellite moves away from the Moon. When the satellite reaches point $%
1\times 10^{7}$, the increase in the time difference gradually slows down.
From Eq. (\ref{trms}), the simplified expression for $-5.68318\times
10^{-7}+(2.97897\times 10^{12})/r^{3}$ shows that the time difference
between the two clocks increases as the satellite moves away from the lunar
surface because the influence of the LTR on clock A on the satellite
gradually decreases. When the satellite moves to an infinite distance from
the Moon, the time difference between the two clocks caused by the LTR is $-5.68318\times
10^{-7}$ $\mu $s/day.

When clock B on the lunar surface is taken to a satellite at a distance $%
1.8\times 10^{6}$ from the Moon's center of mass, and then the satellite
carrying clock A moves away from the Moon, the time difference between the
two clocks caused by the LTR will gradually increase (see the absolute
values represented by the blue dashed line in Fig. 6). From Eq. (\ref{trss}%
), simplified to expression $-5.10797\times 10^{-7}+(2.97897\times
10^{12})/r^{3}$, it can be seen that when the satellite carrying clock A
moves to an infinite distance from the Moon, the time difference between the
two clocks caused by the LTR is $-5.10797\times 10^{-7}$ $\mu $s/day. When
the distance of satellite carrying clock B from the Moon's center of mass is
changed to $1.9\times 10^{6}$, equation $-4.34315\times
10^{-7}+(2.97897\times 10^{12})/r^{3}$, derived from simplifying Eq. (\ref%
{trss}), shows that when the satellite carrying clock A moves to an infinite
distance from the Moon, the time difference between the two clocks due to
the Moon's tidal response is $-4.34315\times 10^{-7}$ $\mu $s/day (the gray
dashed line in Fig. 6). Increasing the distance of clock B from the Moon's
center of mass reduces the influence of the Moon's tidal response on clock
B, indirectly reducing the time difference between the two clocks caused by the LTR.

\section{Conclusions}

In the LCRS, we obtained the FFS between two clocks caused by the LTR at two different
locations on the lunar surface under the Keplerian expansions method. We
found that the magnitude of the FFS caused by the LTR, $10^{-18}$, is close
to the instability and uncertainty achievable by state-of-the-art clocks.
This means that the impact of LTR needs to be taken into account when the
time of multiple devices near the Moon needs to be synchronized to a very
high accuracy.

We specifically analyze the FFS caused by the LTR between clocks at two
different locations on the lunar surface. When the two clocks are fixed at
the same latitude on the lunar surface and one clock is fixed at zero
degrees longitude, the FFS is symmetrically distributed along the $90$%
-degree longitude. When the longitude difference decreases or increases, the
FFS will decrease. Specifically, when the longitude difference between the
two clocks is fixed and the latitude of the two clocks is increased, the FFS
will decrease. When the latitude increases to $90$ degrees, the positions of
the two clocks coincide, and there is no FFS between the two clocks. When
the two clocks are fixed at the same longitude on the lunar surface and a
clock is fixed at the zero degrees latitude, the FFS will increase with the
increase of the latitude difference. Specifically, the FFS decreases when
the latitude difference between the two clocks is fixed and the longitudes
of the two clocks are close to $90$ degrees.

Using the Keplerian expansions method, we also consider the effect of the
Moon's elliptical orbit. The Moon's motion alters both the magnitude and
phase of the FFS caused by the LTR for two clocks that differ only in
longitude. However, when the two clocks differ only in latitude, the Moon's
motion alters only the magnitude of the FFS caused by the LTR. The position of clocks on the
lunar surface and the motion of the Moon produce FFS between the clocks caused by the LTR.

We estimated the time difference between a clock at the highest point on the
lunar surface and a clock at the lowest point on the lunar surface, caused
by the LTR, to be $6.43148\times 10^{-7}$ $\mu $s/day. LTR can affect not
only clocks on the lunar surface but also clocks far away from the lunar
surface. We compared a clock fixed on the lunar surface with a clock placed
on a satellite far from the Moon. As the satellite's clock moves farther
away from the Moon, the time difference between the two clocks caused by LTR
gradually increases. Specifically, the time difference between the two
clocks caused by LTR when the satellite reaches Earth is $-5.68\times
10^{-7} $ $\mu $s/day, and this time difference can be considered the time
difference between the lunar surface and the Earth's surface caused by LTR.
Studying the FFS and time difference caused by the LTR is of great
significance for synchronizing clocks near the Moon and constructing lunar
time.

\section*{Acknowledgments}

We thank the anonymous reviewer for constructive comments that significantly improved this paper.
This work is supported by National Key Research and Development
Program of China with No. 2024YFC2207300, National Natural Science Foundation of China
(NSFC) with Grant No. 12375057, and the Fundamental Research Funds
for the Central Universities, China University of Geosciences (Wuhan).

\end{document}